\documentclass[12pt,oneside,reqno]{amsart}
\usepackage{graphicx}
\usepackage{dcolumn}
\usepackage{amsmath,amsthm,amssymb}
\usepackage{todonotes}
\usepackage{caption}

\theoremstyle{remark}

\theoremstyle{remark}

\begin{document}
\title[Yang-Mills on extremal Reissner-Nordstr\"om]{A Yang-Mills field on the extremal Reissner-Nordstr\"om black hole}

\author{Piotr Bizo\'n}
\address{Institute of Physics, Jagiellonian
University, Krak\'ow, Poland\\ and
Max Planck Institute for Gravitational Physics (Albert Einstein Institute),
Potsdam, Germany}
\email{piotr.bizon@aei.mpg.de}

\author{Micha\l{} Kahl}
\address{Institute of Physics, Jagiellonian
University, Krak\'ow, Poland}
\email{michal.kahl@uj.edu.pl}
\date{\today}%
\begin{abstract}
We consider a spherically symmetric (magnetic) $SU(2)$ Yang-Mills  field propagating on the exterior of the extremal Reissner-Nordstr\"om  black hole. Taking advantage of the conformal symmetry, we reduce the problem to the study of the Yang-Mills equation in a geodesically complete  spacetime with two asymptotically flat ends. We prove the existence of infinitely many static solutions (two of which are found in closed form)  and determine the spectrum of their linear perturbations and quasinormal modes. Finally, using the hyperboloidal approach to the initial value problem, we describe the process of relaxation to the static endstates of evolution for various initial data.
\end{abstract}

\maketitle

\noindent

\section{Introduction and setup}
The global dynamics of a Yang-Mills  field propagating in a four-dimensional Minkowski spacetime is well understood: all  solutions starting from smooth initial data at $t=0$  remain smooth for all times \cite{em} and decay to zero as $t\rightarrow \pm\infty$ \cite{ch,bcr}. The global-in-time regularity holds true in any globally hyperbolic four-dimensional curved spacetime \cite{cs}, however the phase portrait can be richer due to existence of nontrivial stationary solutions which play the role unstable attractors, as was shown for the Schwarzschild background~\cite{brz}.

     In this paper we consider the evolution of a Yang-Mills field on the exterior of the extremal Reissner-Nordstr\"om  black hole. Our study is motivated by the expectation that an interplay between  the conformal structure of the extremal Reissner-Nordstr\"om black hole and the conformal invariance of the Yang-Mills equations may lead to an interesting behaviour of solutions. This is a continuation of our studies of how the dissipation-by-dispersion phenomena, responsible for the relaxation to a stationary equilibrium in extended Hamiltonian systems, depend on the geometry of the underlying spacetime \cite{bk, bm}.

 The exterior (i.e., the domain of outer communication)
 of the extremal Reissner-Nordstr\"om black hole is a globally hyperbolic static spacetime $(\mathcal{M},\hat g)$ whose metric, in coordinates $t\in \mathbb{R}, r>M,(\vartheta,\varphi)\in S^2$, reads
\begin{equation}\label{rn}
  \hat g=-\left(1-\frac{M}{r}\right)^{2} dt^2+\left(1-\frac{M}{r}\right)^{-2} dr^2 + r^2 (d\vartheta^2+\sin^2{\!\vartheta}\, d\varphi^2)\,,
\end{equation}
where $M$ is  positive constant. The metric $\hat g$ is a spherically symmetric solution of the Einstein-Maxwell equations with mass $M$ and charge $Q=\sqrt{M}$. In order to better see the global properties of this spacetime, it is convenient to use dimensionless  variables $(\tau,x)\in \mathbb{R}^2$:
  \begin{equation}
  \tau=\frac{t}{4M}, \quad \mbox{and} \quad  x=\ln\left(\frac{r}{M}-1\right)\,,
  \end{equation}
in terms of which the metric \eqref{rn} takes the form
  \begin{equation}\label{rnx}
  \hat g=\frac{16 M^2 }{(1+e^{-x})^2} \; g\,,
\end{equation}
where
\begin{equation}\label{sigma}
  g= -d\tau^2 + C^4 \, (dx^2 + d\vartheta^2+\sin^2{\!\vartheta}\, d\varphi^2)\,.
\end{equation}
Hereafter, for typographical convenience we use the abbreviation
  $C=C(x)=\cosh{\frac{x}{2}}$. We note that the Ricci scalar vanishes both for $\hat g$ and $g$.
In contrast to $(\mathcal{M}, \hat g)$, the spacetime $(\mathcal{M}, g)$ is geodesically complete. It has two asymptotically flat ends at $x=\pm \infty$ (see Fig.~1). Asymptotic flatness is easily seen in terms of the coordinate $\rho=C^2(x)$ for which we have
 \begin{equation}\label{RNr}
 g= -d\tau^2 + \left(1-\frac{1}{\rho}\right)^{-1}  d\rho^2 + \rho^2 (d\vartheta^2+\sin^2{\!\vartheta}\, d\varphi^2)\,.
 \end{equation}
 \begin{figure}[h]
  \begin{center}
    \includegraphics[width=0.35\textwidth]{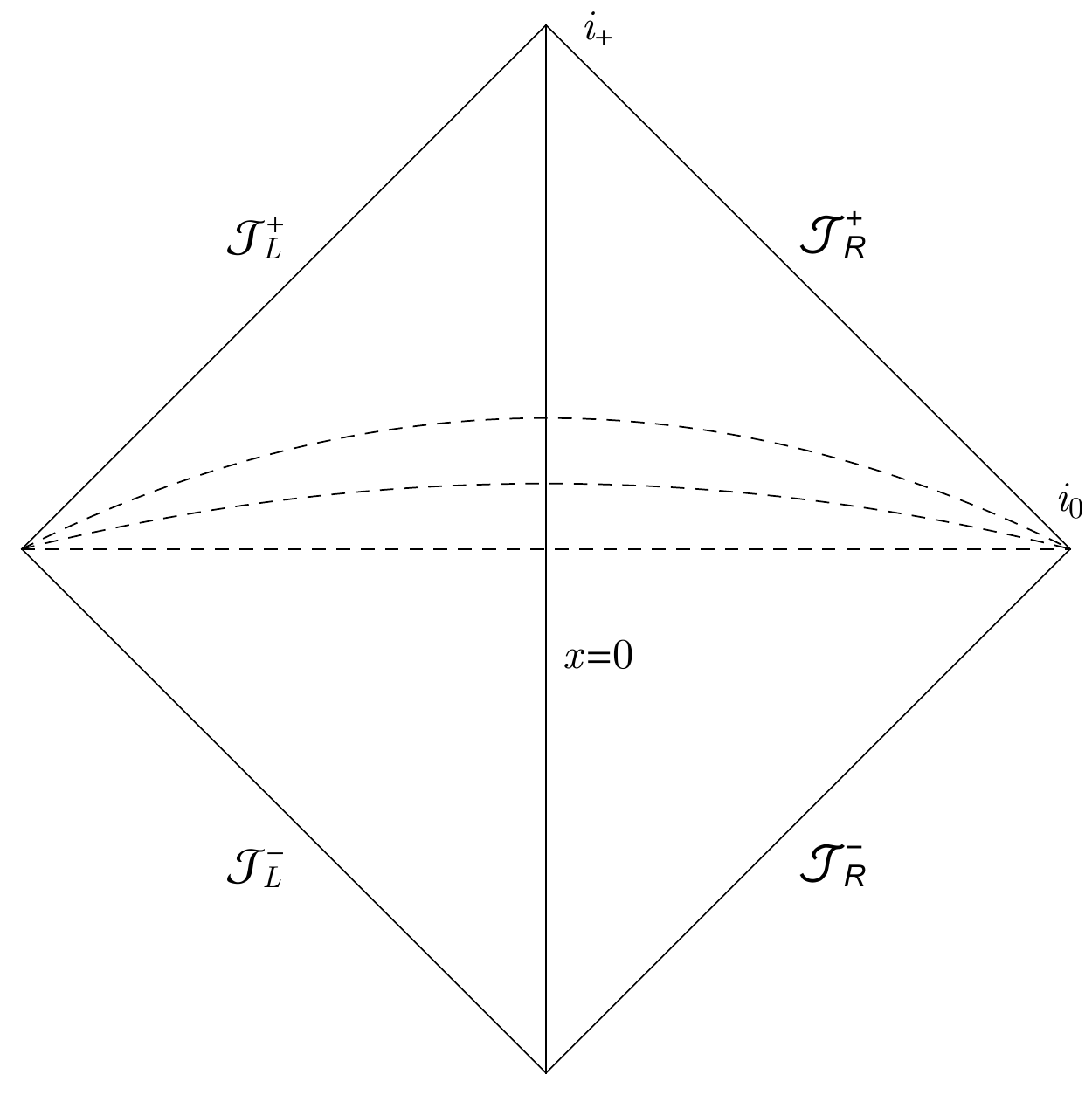}
    \caption{{\small Penrose diagram for $(\mathcal{M},g)$.}}
    \label{fig1}
  \end{center}
\end{figure}

Note that the reflection $x\mapsto -x$ is the isometry of the  metric $g$ but only the conformal isometry for the metric $\hat g$ \cite{ct}. On $(\mathcal{M},g)$ the reflection interchanges the 'left' and the 'right'  future null infinities, ${\mathcal J}_L^+$ and ${\mathcal J}_R^+$, while on $(\mathcal{M},\hat g)$ it interchanges the event  horizon and ${\mathcal J}_R^+$.

We consider an $SU(2)$ Yang-Mills   field propagating in the spacetime $(\mathcal{M},g)$. The gauge potential $A_{\mu}=A_{\mu}^a \tau_a$ takes values in the Lie algebra $su(2)$, where the generators $\tau_a$   satisfy $[\tau_a,\tau_b]=i \epsilon_{abc} \tau_c$. In terms of the Yang-Mills field strength $F_{\mu\nu}=\nabla_{\mu} A_{\nu} - \nabla_{\nu} A_{\mu} +[A_{\mu}, A_{\nu}]$, the lagrangian density reads
\begin{equation}\label{action}
  \mathcal{L} = \mathrm{Tr}\left(F_{\alpha\beta} F_{\mu\nu} g^{\alpha\mu} g^{\beta\nu}\right)\, \sqrt{-\text{det}(g_{\mu\nu})}.
\end{equation}
In four dimensions the quantity $g^{\alpha\mu} g^{\beta\nu}\, \sqrt{-\text{det}(g_{\mu\nu})}$ is invariant under a conformal transformation $g_{\mu\nu} \mapsto\Omega^2 g_{\mu\nu}$, hence if $A_{\mu}$ solves the Yang-Mills equations in one metric, so does it in any conformally related metric. Taking advantage of this conformal invariance of the Yang-Mills equations, in the following we consider only the spacetime $(\mathcal{M},g)$.

For the  Yang-Mills potential we assume the  spherically symmetric purely magnetic ansatz
\begin{equation}\label{A}
  A = W(\tau,x)\, \omega +\tau_3 \cos{\vartheta} d\varphi\,,\quad\mbox{where} \quad \omega=\tau_1 d\vartheta+\tau_2 \sin{\vartheta} \,d\varphi\,,
\end{equation}
 which gives
\begin{equation}\label{F}
  F=\partial_{\tau} W d\tau\wedge \omega + \partial_x W dx\wedge\omega -(1-W^2)\, \tau_3 \,d\vartheta \wedge \sin{\vartheta} \,d\varphi\,.
\end{equation}
Note that the vacuum is given by $W=\pm 1$, while $W=0$ corresponds to the magnetic monopole with unit charge.
Inserting the ansatz \eqref{F} into \eqref{action} we get the reduced lagrangian density
\begin{equation}\label{red_action}
 \mathcal{L}=-\frac{1}{2} \left(\partial_{\tau} W\right)^2 C^2+\frac{1}{2} \left(\partial_{x} W\right)^2 C^{-2}+\frac{1}{4} \left(1-W^2\right)^2 C^{-2}\,,
\end{equation}
and the corresponding Euler-Lagrange equation
\begin{equation}\label{ymx}
 \partial_{\tau\tau} W= C^{-2} \partial_x \left(C^{-2}  \partial_x W\right) + C^{-4} W (1-W^2)\,.
\end{equation}
We know from Chru\'sciel and Shatah \cite{cs} that solutions of Eq.\eqref{ymx} starting at $\tau=0$ from smooth initial data remain smooth for all future times. The goal of this paper is to describe their asymptotic behaviour for $\tau\rightarrow \infty$. For physical reasons our analysis is restricted to solutions with finite (conserved) energy
\begin{equation}\label{energy}
  E=\frac{1}{2} \int_{-\infty}^{\infty} \left[C^{2} (\partial_{\tau} W)^2
+ C^{-2}\left( (\partial_x W)^2+\frac{(1-W^2)^2}{2}\right)\right] dx < \infty\,.
\end{equation}
 Due to dissipation of energy by dispersion, such solutions are expected to settle down to critical points of the potential energy, i.e. static
 solutions of Eq.\eqref{ymx}.

\section{Static solutions}
\noindent Time-independent solutions $W=W(x)$ of Eq.\eqref{ymx} satisfy the ordinary differential equation
\begin{equation}\label{ode}
  W''-\tanh\left(\frac{x}{2}\right) W'+W (1-W^2)=0\,.
\end{equation}
We claim that, besides the constant solution $W_0=1$, Eq.\eqref{ode} has a countable family of smooth finite energy solutions $W_n(x)$ ($n\in \mathbb{N})$ with the following properties (which, not very surprisingly, bear remarkable similarities to the Bartnik-McKinnon solutions of  Einstein-Yang-Mills equations \cite{bmk}):
\begin{itemize}
\item $W_n(x)$ has $n$ zeros,
\item $|W_n(x)|<1$ for all finite $x$ and $\lim_{|x|\rightarrow\infty} |W_n(x)| = 1$,
\item $W_n(x)$ is an even (resp. odd) function for even (resp. odd) $n$,
\item As $n\rightarrow\infty$, $W_n(x)$ tend pointwise to $W_{\infty}=0$ for any finite $x$.
\end{itemize}
The proof of existence of solutions $W_n$ and their properties is a straightforward adaptation of the proof given in \cite{hm} in the case of harmonic maps between 3-spheres (which satisfy the same equation as \eqref{ode} with the nonlinearity $\sin(2W)$ instead of $W(1-W^2)$). Key to the proof is the fact that Eq.\eqref{ode} is asymptotically autonomous with the limiting equations for $x\rightarrow\pm\infty$,
$ W''\mp W'+W (1-W^2)=0$,
having saddle points at $W=\pm 1$ and a spiral at $W=0$ (stable at $-\infty$ and unstable at $+\infty$). Using a shooting method one can show that there exist infinitely many homoclinic and heteroclinic orbits connecting the saddle points. The solutions are parametrized by the coefficients of the stable directions  of the saddle points
\begin{equation}\label{expinfty}
W_n(x)=1-a_n e^{-x}+\mathcal{O}(e^{-3x}).
\end{equation}
We refer the interested reader to \cite{hm} for the details. Note that due to the reflection symmetry $W\mapsto -W$, each solution $W_n$ has a copy $-W_n$.
We adopt the convention that $W_n(\infty)=1$. The  parameters and energies of the first few solutions are given in Table~1 and their profiles  are depicted in Fig.~2.
\begin{table}[!h]
  \centering
  \begin{tabular}{|c|cccccc|}\hline
    \noalign{\smallskip}
     $\,\,$n$\,\,$ & $1$ & $2$ & $3$ & $4$ & $5$ & $6$ \\
    \noalign{\smallskip}\hline\noalign{\smallskip}
    $\,\,a_{n}\,\,$ & $\,\,2.0\,\,$  & $\,\,15.798\,\,$  & $\,\,101.108\,\,$  & $\,\,624.538\,\,$  & $\,\,3835.14\,\,$  & $\,\,23528\,\,$ \\ 
    $\,\,E_{n}\,\,$ &  $\,\,0.8\,\,$   & $\,\,0.9664\,\,$  & $\,\,0.9945\,\,$  & $\,\,0.9991\,\,$  & $\,\,0.99985\,\,$  & $\,\,0.999976\,\,$  \\[1ex]
    \hline
  \end{tabular}
  \vspace{0.2cm}
  \caption{{\small The parameters of the first few static solutions $W_n(x)$.}}
  \label{tab:static.params}
\end{table}
\vskip 0.2cm
\begin{figure}[h]
\begin{center}
\includegraphics[width=0.49\textwidth]{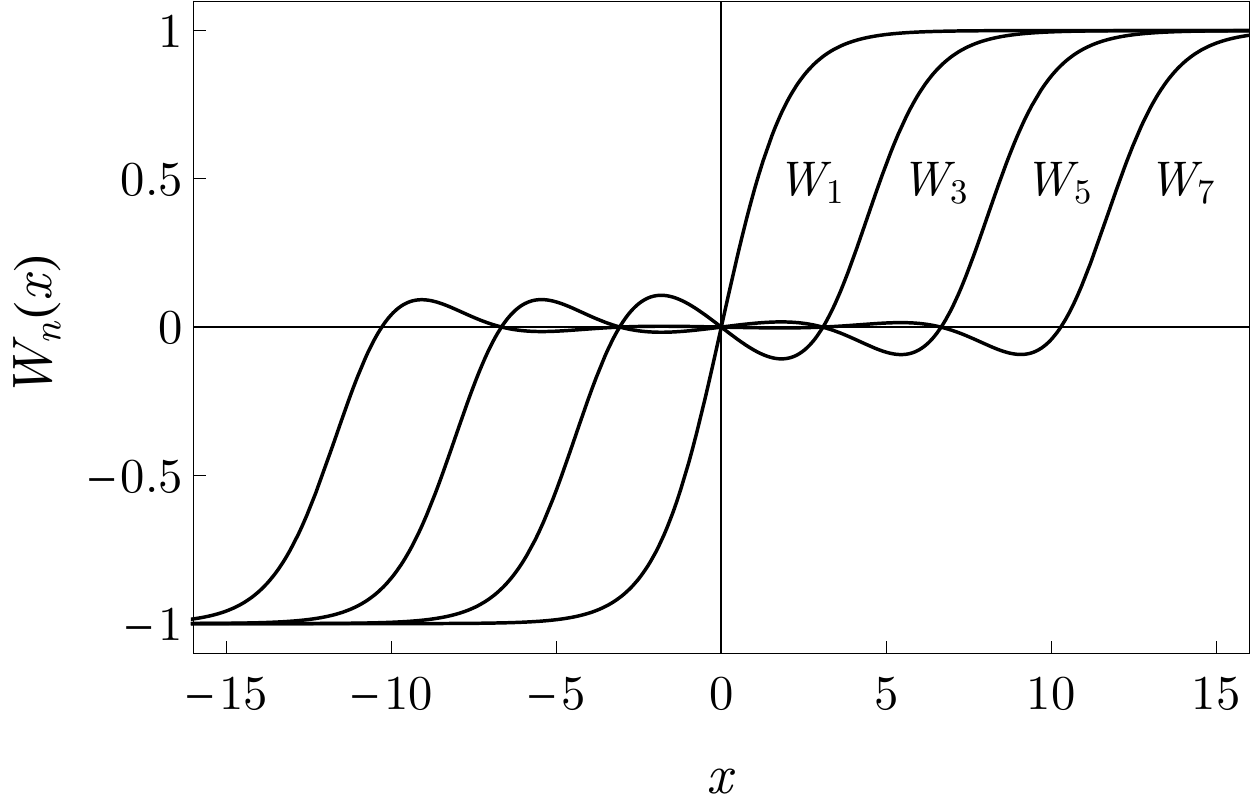}
\includegraphics[width=0.49\textwidth]{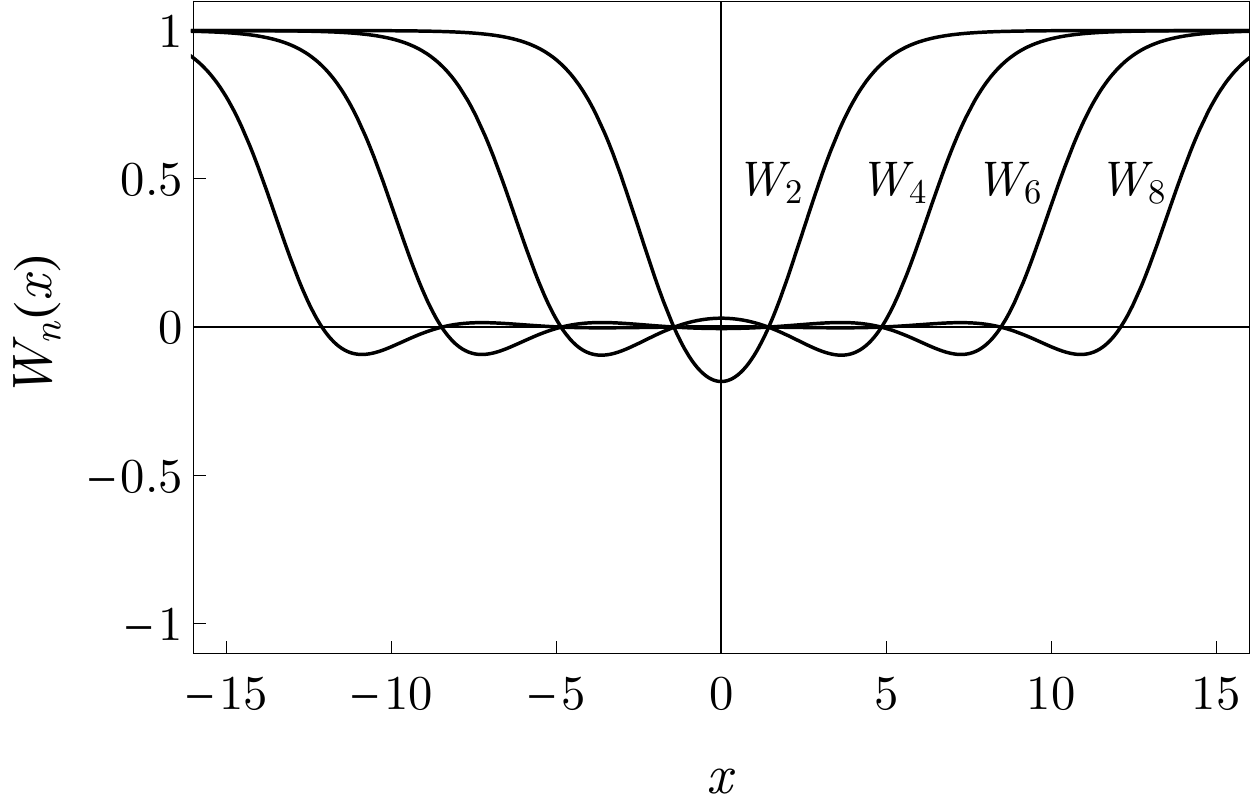}
\caption{{\small The first four odd  and even static solutions.}}
\label{fig2}
\end{center}
\end{figure}
 Experimenting with Maple, we found the first two nontrivial solutions in closed form
\begin{equation}\label{exact}
  W_1(x)=\tanh\left(\frac{x}{2}\right),\qquad W_2(x)=\frac{2\cosh{x}-2-\sqrt{6}}{2\cosh{x}+4+3\sqrt{6}}\,.
\end{equation}
\vskip 0.3cm
 We turn now to the linear stability analysis of the static solutions $W_n(x)$. This analysis is essential in understanding the role  these solutions may play in the evolution.
 Following the standard procedure, we substitute $W(\tau,x)=W_n(x)+\delta W(\tau,x)$ into Eq.\eqref{ymx}, linearize and separate the time dependence $\delta W(\tau,x)=e^{\lambda \tau} v(x)$. This yields  the eigenvalue problem\footnote{We point out in passing that in terms of the variable $y=\frac{1}{2} (\sinh{x}+x)$, defined by $dy/dx=C^2(x)$, Eq.\eqref{pert} takes the form of the one-dimensional Schr\"odinger equation
$-\frac{d^2 v}{dy^2}+V_n(x(y)) v(y) =-\lambda^2 v(y)$, however this form
 is not very helpful because the function $x(y)$ is given only implicitly.}
\begin{equation}\label{pert}
 L_n\, v:= -\frac{1}{C^2}\,\frac{d}{dx}\left(\frac{1}{C^2}\,\frac{d v}{dx}\right)+\frac{3 W_n^2(x)-1}{C^4(x)}\,v=-\lambda^2\,v \,.
\end{equation}
 Note that due to the reflection symmetry $x\rightarrow -x$,  the eigenfunctions are alternately even and odd.  We denote them by $v^{(n)}_k(x)$ and the corresponding eigenvalues by $-(\lambda_k^{(n)})^2$ ($k\in \mathbb{N}$).
 We claim that the operator $L_n$ has exactly~$n$ negative eigenvalues. For $n=0$ this is evident because the potential in \eqref{pert} is everywhere positive. For $n=\infty$ this follows from the fact that the $\lambda=0$ solution is oscillating at infinity. For any finite $n\geq 1$, one can obtain a lower bound as follows. Consider the function $u_n(x):=W'_n(x)$. Differentiating Eq.\eqref{ode}, one finds that
 \begin{equation}\label{U}
   \tilde L_n\, u_n=0\,, \quad\mbox{where}\quad \tilde L_n=L_n + \frac{1}{2 C^6(x)}\,.
 \end{equation}
 By construction $u_n(x)$ has $(n-1)$ zeros, hence from the Sturm oscillation theorem it follows that the operator $\tilde L_n$ has exactly $(n-1)$ negative eigenvalues. Consequently, the operator~$L_n$, which is the exponentially localized negative perturbation of $\tilde L_n$,  has a least $(n-1)+1=n$ negative eigenvalues (where '$+1$' stands for the zero eigenvalue of $\tilde L_n$ going negative). Numerics (see Table~2) shows that this lower bound is sharp  but it seems hard to prove this fact rigorously for general~$n$. In the case $n=1$,  non-existence of the second negative eigenvalue follows by the Sturm oscillation theorem from an easy to prove fact that the odd solution  solution of Eq.\eqref{pert} with $\lambda=0$ is monotone.
\begin{table}[!h]
  \centering
  \begin{tabular}{|c|cccc|}\hline
    \noalign{\smallskip}
     $\,\,$n$\,\,$ & $\lambda^{(n)}_1$ & $\lambda^{(n)}_2$ & $\lambda^{(n)}_3$ & $\lambda^{(n)}_4$ \\
    \noalign{\smallskip}\hline\noalign{\smallskip}
    $\,\,$1$\,\,$ & $\,\,0.54089 \,\,$ & $\,\, \,\,$ & $\,\,\,\,$ & $\,\,\,\,$ \\
    $\,\,$2$\,\,$ & $\,\,0.69937 \,\,$ & $\,\,0.17161 \,\,$ & $\,\,\,\,$ & $\,\,\,\,$ \\
    $\,\,$3$\,\,$ & $\,\,0.72553 \,\,$ & $\,\,0.21010 \,\,$ & $\,\,0.033792\,\,$ & $\,\,\,\,$ \\
    $\,\,$4$\,\,$ & $\,\,0.72934 \,\,$ & $\,\,0.21772 \,\,$ & $\,\,0.040983\,\,$ & $\,\,0.0057005\,\,$ \\
    $\,\,\infty\,\,$ & $\,\,0.73015 \,\,$ & $\,\,0.21884 \,\,$ & $\,\,0.042775\,\,$ & $\,\,0.0072103\,\,$ \\[1ex]
    \hline
  \end{tabular}
  \vspace{0.3cm}
  \caption{{\small Lyapunov exponents of the unstable modes of $W_n$.}}
  \label{tab:unstable}
\end{table}
\section{Hyperboloidal formulation}
 We will use the method of hyperboloidal foliations
and Scri-fixing as developed by Zengino\u{g}lu \cite{anil1} on the basis of concepts introduced by Friedrich \cite{f}. To implement this method we define a new time coordinate
\begin{equation}\label{s}
  s=\tau - \frac{1}{2} \left(\cosh{x}+\ln(2\cosh{x})\right)
\end{equation}
and foliate the spacetime by hyperboloidal hypersurfaces $\Sigma_s$ of constant $s$. These are spacelike hypersurfaces that approach the 'left' future null infinity along  outgoing null cones of constant advanced time $v=\tau + \frac{1}{2} (\sinh{x}+x)$ and the 'right' future null infinity along  outgoing null cones of constant retarded time $u=\tau - \frac{1}{2} (\sinh{x}+x)$. To see this, note that
\begin{equation*}\label{n}
  (\partial_{\alpha} s)(\partial_{\beta} s) g^{\alpha\beta} = -\frac{1}{\cosh^2{\!x}} \rightarrow 0 \quad\mbox{as}\quad |x|\rightarrow \infty\,,
\end{equation*}
and
\begin{eqnarray*}\label{u}
v_{\vert_{\Sigma_s}}&=&s+\frac{1}{2} (\cosh{x}+\ln(2\cosh{x}))+ \frac{1}{2} (\sinh{x}+x)\rightarrow s \quad \mbox{as} \quad x\rightarrow-\infty\,,\\
u_{\vert_{\Sigma_s}}&=&s+\frac{1}{2} (\cosh{x}+\ln(2\cosh{x}))- \frac{1}{2} (\sinh{x}+x)\rightarrow s \quad \mbox{as} \quad x\rightarrow\infty\,.
\end{eqnarray*}
There is plenty of freedom in choosing a hyperboloidal foliation; the particular choice \eqref{s} is motivated by computational convenience.
In terms of the coordinates $(s,x)$, Eq.\eqref{ymx} takes the  form\footnote{We slightly abuse notation and use the same letter for the function $W(\tau,x)$ and any other function obtained from it by changing the variables.}
\begin{equation}\label{eqs}
 \frac{C^2}{\cosh^2{\!x}}\partial_{ss} W +2 \tanh{x}\, \partial_{x s} W + \frac{1}{\cosh^{2}{\!x}} \,\partial_s W = \partial_x\left(C^{-2}\partial_x W\right) + C^{-2} W (1-W^2)\,.
\end{equation}

 Next, we compactify the real line $-\infty<x<\infty$  to the interval $[-1,1]$ by  the coordinate transformation $z=\tanh{\frac{x}{2}}$. This fixes ${\mathcal J}_L^+$ at $z=-1$ and ${\mathcal J}_R^+$ at $z=1$. Eq.\eqref{eqs} now becomes
  \begin{equation}\label{eqz}
 \frac{1}{(1+z^2)^2}\, \partial_{ss} W +\frac{2z}{1+z^2} \,\partial_{z s} W + \frac{1-z^2}{(1+z^2)^2} \,\partial_s W=\partial_z\left(\frac{(1-z^2)^2}{4}\, \partial_z W\right) + W (1-W^2)\,.
\end{equation}
Multiplying this equation  by $\partial_s W$ we obtain the local conservation law
\begin{equation}\label{cons}
 \partial_s e +\partial_z f=0\,,
\end{equation}
where
\begin{eqnarray*}\label{cons2}
e(s,z)&=& \frac{1}{(1+z^2)^2}\, (\partial_s W)^2 + \frac{(1-z^2)^2}{4}\,(\partial_z W)^2
  +\frac{(1-W^2)^2}{2}\,, \\
f(s,z)&=&\frac{(1-z^2)^2}{2}\,\partial_s W \partial_z W-\frac{2z}{1+z^2}\,(\partial_s W)^2\,.
\end{eqnarray*}
Integrating \eqref{cons} over a hypersurface $\Sigma_s$ we get the energy balance
\begin{equation}\label{flux}
  \frac{d\mathcal{E}}{ds} =-(\partial_s W(s,1))^2-(\partial_s W(s,-1))^2\,,
\end{equation}
where
\begin{equation}\label{bondi}
  \mathcal{E}(s)=\int_{-1}^{1} e(s,z) dz
\end{equation}
is the Bondi-type energy. The formula \eqref{flux} expresses the radiative loss of energy through the future null infinities. Since the energy $\mathcal{E}(s)$ is positive and monotone decreasing, it has a nonnegative limit for $s\rightarrow \infty$.  For this reason the hyperboloidal formulation is very natural in analyzing relaxation processes that are due to the dispersive dissipation of energy. In the remainder of the paper we describe in detail the convergence to one of the static solutions $W_n(z)$. We focus our attention on the first two static solutions  $W_0$ and $W_1$ because, as follows from the linear stability analysis in section~2, only these solutions may participate in the evolution of generic and codimension-one initial data.

We return now to  the linear perturbation analysis and compute the quasinormal modes for $W_0$ and $W_1$. Substituting
$$
W(s,z)=W_n(z)+(1+z^2)^{\lambda/2} e^{\lambda s} u(z)
$$ into \eqref{eqz} and linearizing, we obtain the quadratic eigenvalue problem (the purpose of the factor $(1+z^2)^{\lambda/2}$ is to simplify the resulting equation)
\begin{equation}\label{pert2}
  (1-z^2)^2 u''+2 z \left(\lambda(z^2-3)+2z^2+2\right) u'+ \left(\lambda^2 (z^2-4)+3\lambda (z^2-1) + 4-12 W_n^2\right) u=0\,.
\end{equation}
 We shall compute the quasinormal modes for the first two static solutions $W_0=1$ and $W_1=z$  using  Leaver's method\footnote{An alternative (black box) method is to exploit  the fact that Eq.\eqref{pert2} for $n=0,1$ has the form of a double confluent Heun equation and get Maple to do the rest of the job \cite{bk, fs}, however we prefer a more transparent approach.} \cite{leaver}.
  We seek solutions  in the form of a power series  around the ordinary point $z=0$
\begin{equation}\label{series}
u(z)=\sum_{j=0} a_j z^j\,.
 \end{equation}
This series  converges for $|z|<1$. The discrete values of $\lambda$ for which the function defined by the series \eqref{series} is \emph{analytic} at $z=\pm 1$ correspond to the eigenvalues (for real $\lambda>0$) and quasinormal modes (for $\Re(\lambda)<0$). To find those values we need to determine the asymptotics of the coefficients $a_j$ for large $j$.
To this end we substitute the series \eqref{series} into Eq.\eqref{pert2} for $n=0,1$ and  get the three-term recurrence relation
\begin{align}\label{rec}
 & \alpha_0\, a_2 + \beta_0\, a_0=0\,, \nonumber\\
 & \alpha_1\, a_3 + \beta_1\, a_1=0\,,\nonumber \\
 &  \alpha_j\, a_{j+2} +\beta_j\, a_{j} +\gamma_j\, a_{j-2}=0\,, \quad j\geq 2
\end{align}
with
\begin{eqnarray*}
\alpha_j&=&j^2+3j+2\,,\\
\beta_j&=&-2 j^2 -(6\lambda+2) j -4 \lambda^2 - 3\lambda +4-12 b_0^2\,,\\
\gamma_j&=& j^2+(2\lambda-1) j +\lambda^2-\lambda-2-12b_1^2\,,
\end{eqnarray*}
where $b_0$ and $b_1$ are the first two coefficients of the Taylor expansion of $W_n(z)$ around $z=0$. For $n=0$ we have $b_0=1, b_1=0$ and for $n=1$ we have $b_0=0,b_1=1$.
Even modes satisfy the boundary condition $u(0)=1, u'(0)=0$, hence $a_0=1$ and $a_1=0$, while for odd modes we have $u(0)=0, u'(0)=1$, hence $a_0=0$ and $a_1=1$.

The recurrence relation \eqref{rec} has two linearly independent asymptotic solutions for $j\rightarrow \infty$
(in the theory of finite difference equations, such solutions are called Birkhoff's solutions, see Chapter 8.6 in \cite{pin})
\begin{equation}\label{birk}
  a_j^{\pm} \sim j^{\beta} \exp(\pm 2 \sqrt{\lambda}\sqrt{j})\,,\quad\mbox{where}\quad \beta=\frac{\lambda}{2}-\frac{3}{4}\,,
\end{equation}
(the leading order behaviour does not depend on $n$), thus asymptotically
\begin{equation}\label{c12}
a_j \sim c_{+}(\lambda) a_j^{+} + c_{-}(\lambda) a_j^{-}\,.
 \end{equation}
 The series \eqref{series}  converges at $z=\pm 1$ if and only if the coefficient of an exponentially growing term  in \eqref{c12} vanishes. The corresponding solution is then called a minimal solution of the recurrence relation and denoted $a_j^{min}$. To find the minimal solution we define
\begin{equation*}\label{ab}
  A_j=\frac{\beta_j}{\alpha_j},\qquad  B_j=\frac{\gamma_j}{\alpha_j},\qquad  r_j=\frac{a_{j}}{a_{j-2}}\,,
\end{equation*}
and rewrite \eqref{rec} in the form
\begin{equation}\label{cf0}
  r_j=-\frac{B_j}{A_j+r_{j+2}}\,.
\end{equation}
Iterating this formula we get the continued fraction representation
\begin{equation}\label{cf}
  r_j=-\frac{B_j}{A_j-}\,\frac{B_{j+2}}{A_{j+2}-}\,\frac{B_{j+4}}{A_{j+4}-}\,\dots\,.
\end{equation}
According to Pincherle's theorem \cite{pin}, this continued fraction converges if and only if the recurrence relation \eqref{rec} has a minimal solution $a_j^{min}$ and then $r_j=a_{j}^{min}/a_{j-2}^{min}$ for each $j\geq 2$.
Using Pincherle's theorem  we obtain the following quantization conditions
\begin{equation}\label{quant0}
  \frac{a_2}{a_0}=-\frac{B_0}{A_0-}\,\frac{B_{2}}{A_{2}-}\,\frac{B_{4}}{A_{4}-}\,\dots\ \quad\mbox{for even modes}\,,
\end{equation}
\begin{equation}\label{quant1}
  \frac{a_3}{a_1}=-\frac{B_1}{A_1-}\,\frac{B_{3}}{A_{3}-}\,\frac{B_{5}}{A_{5}-}\,\dots\ \quad\mbox{for odd modes}\,.
\end{equation}
The continued fractions above can be computed to any desired precision by downward recursion starting from some large $j_{max}$ and an arbitrary initial value $r_{j_{max}}$. Finally, the roots of the quantization conditions \eqref{quant0} and \eqref{quant1} are determined numerically (see Table~3).
\begin{table}[!h]
  \centering
  \begin{tabular}{|c|ccc|}
    \noalign{\smallskip}\hline\noalign{\smallskip}
    $\,\,W_{0}\,\,$ & $\,\,-0.33756 \pm 1.34121 i \,\,$ & $\,\, -1.04619 \pm 1.21202 i \,\,$ & $\,\, -1.85831 \pm 0.99066 i \,\,$  \\
    $\,\,W_{1}\,\,$ & $\,\, -0.39316 \pm 0.08844 i \,\,$ & $\,\, -0.26656 \pm 0.45639 i \,\,$ & $\,\, -1.39277 \pm 0.15802 i \,\,$  \\[1ex]
    \hline
  \end{tabular}
  \vspace{0.3cm}
  \caption{{\small The  three least damped quasinormal frequencies for $\!W_0\!$ and $\!W_1\!$ (the corresponding modes are alternately even and odd).}}
  \label{tab:qnm}
\end{table}
\section{Numerical results}
Following \cite{anil2} and \cite{bk}, we define the auxiliary variables
\begin{equation}\label{aux}
  \Psi=\frac{1}{2} \partial_z W\,,\quad\mbox{and}\quad \Pi=\frac{1}{(1+z^2)^2}\, \partial_s W + \frac{z}{1+z^2}\, \partial_z W\,,
\end{equation}
  and rewrite Eq.\eqref{eqz} as the first order symmetric hyperbolic system
  \begin{eqnarray}\label{symhyp}
  \partial_s W &=& (1+z^2)^2\, \Pi - 2z(1+z^2)\,\Psi\,,\\
  \partial_s \Psi &=& \partial_z \left(\frac{1}{2} (1+z^2)^2\,\Pi - z (1+z^2)\,\Psi\right)\,,\\
  \partial_s \Pi &=& \partial_z\left(\frac{1}{2} (1+z^2)^2\,\Psi - z (1+z^2)\,\Pi\right)+W(1-W^2)\,.
  \end{eqnarray}
 We solve this system numericaly using the method of lines with an 8th-order finite difference scheme in space and 4th-order Runge-Kutta integration in time. At the boundaries we use one-sided stencils. Note that there are no ingoing characteristics at the boundaries, hence no boundary conditions need, or can, be imposed.

As expected, for generic intial data the solution tends to one of the vacuum solutions $\pm W_0$. For intermediate times we observe ringdown along the fundamental quasinormal mode, while for later times  the  polynomial tail is seen to dominate (see Fig.~3).
\begin{figure}[h!]
  \begin{center}
    \includegraphics[width=0.75\textwidth]{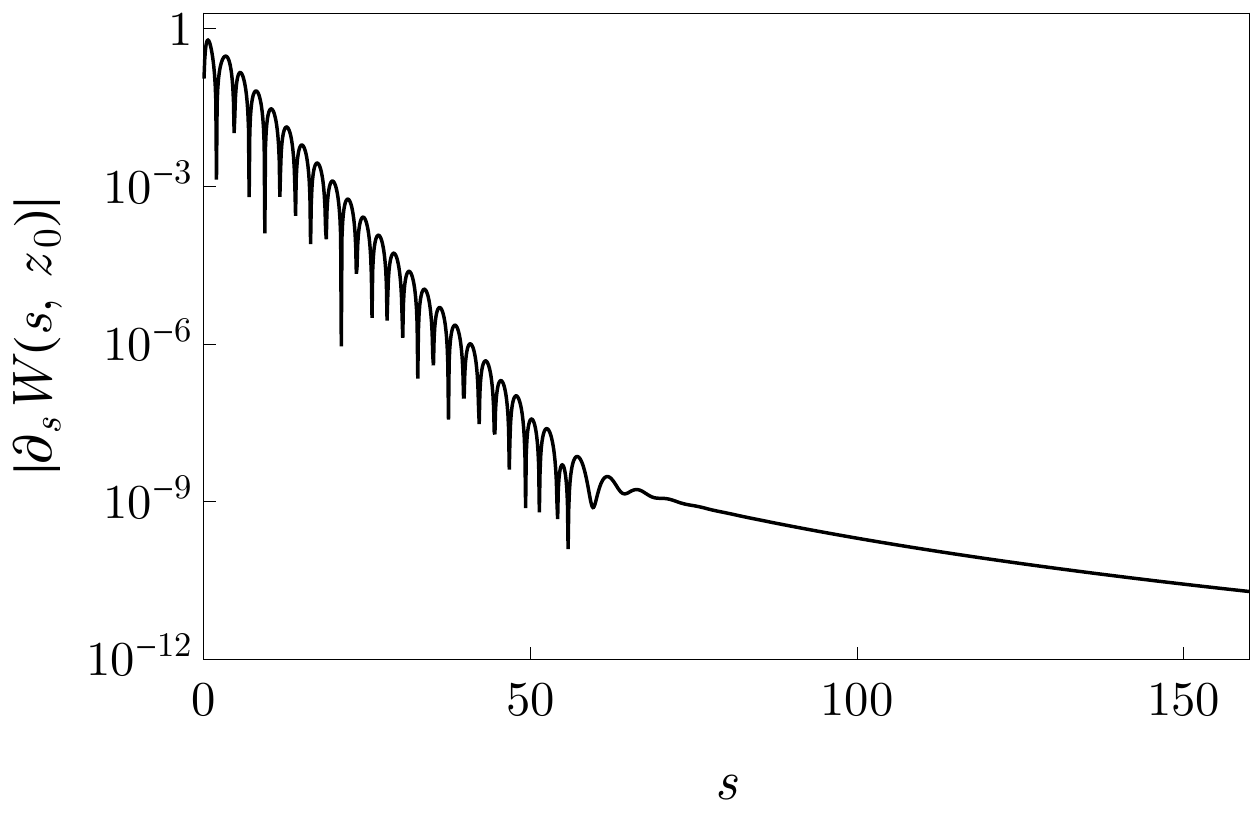}
    \caption{{\small Evolution of time-symmetric initial data $W(0,z)=1+\frac{1}{2}\exp(-2 z^2)$ at a sample point $z=\frac{1}{2}$. The ringdown to $W_0$ for  intermediate times is governed by the  quasinormal mode with frequency $\lambda \approx -0.337+1.341 i$. For late times $|\partial_s W(s,\frac{1}{2})| \sim s^{-5}$. }}
    \label{fig3}
  \end{center}
\end{figure}

If initial data are close to an unstable static solution, then for early times we observe an exponentially fast departure from this solution along its principal unstable mode. This behaviour, in the case of a small perturbation of $W_{\infty}$,  is shown in Fig.~4.

\begin{figure}[h]
  \begin{center}
    \includegraphics[width=0.75\textwidth]{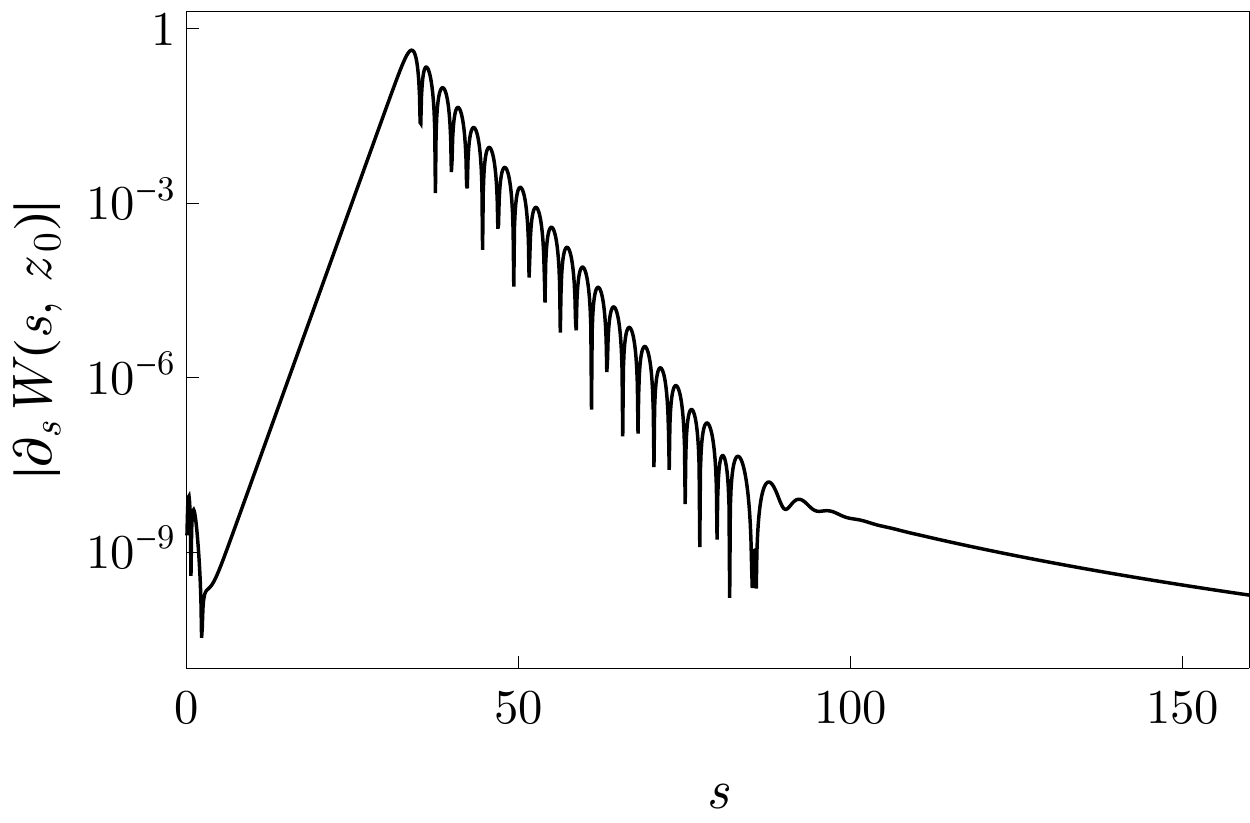}
    \caption{{\small Evolution of time-symmetric initial data $W(0,z)=10^{-8}\,\exp(-16 \text{atanh}^2z)$ at a sample point $z=\frac{1}{2}$. For early times we see the exponential growth along the principal unstable mode around $W_{\infty}$ with the Lyapunov exponent $\lambda^{(\infty)}_1\approx 0.73$. Afterwards, the evolution proceeds as in Fig.~3.}}
    \label{fig4}
  \end{center}
\end{figure}

Since parity is preserved in evolution, solutions starting from odd initial data cannot tend to $W_0$. Generically, odd solutions converge to $W_1$ (whose single unstable mode is even and therefore not excited). If we add a small even admixture of size $\varepsilon$ to odd initial data, then $W_1$ appears as an intermediate attractor with lifetime  $\sim \frac{1}{\lambda^{(1)}_1} \log(\frac{1}{\varepsilon})$, where $\lambda^{(1)}_1\approx 0.54$ is the Lyapunov exponent of the unstable mode of $W_1$  (see Fig.~5).
\begin{figure}[h]
  \begin{center}
    \includegraphics[width=0.75\textwidth]{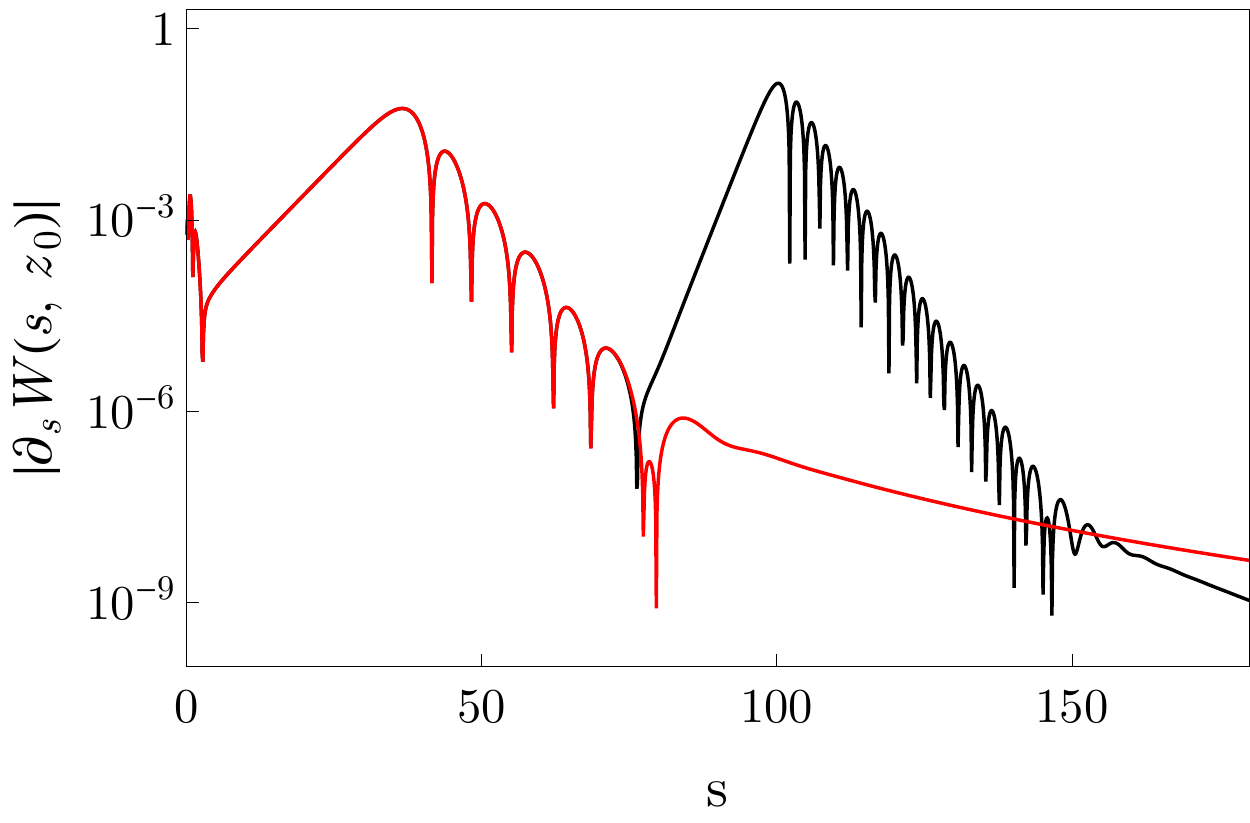}
    \caption{{\small Evolution of time-symmetric initial data $W(0,z)=(0.01\tan(z)-10^{-24})\,\exp(-16 \text{atanh}^2z)$ (black) and $W(0,z)=0.01\tan(z)\,\exp(-16 \text{atanh}^2z)$ (red) at a sample point $z=\frac{1}{2}$. Initially, both solutions evolve together and ring down to $W_1$ along its least damped odd quasinormal mode with frequency $\lambda\approx -0.266+0.456i$. Later, the solution with the small even admixture departs towards $W_0$ along the unstable mode of $W_1$ with the Lyapunov exponent $\lambda^{(1)}_1 \approx 0.54$.}}
    \label{fig5}
  \end{center}
\end{figure}

In all cases, the dynamics in the vicinity of static solutions is in excellent quantitative agreement with the results of linear stability analysis displayed in Tables~2 and 3.

Finally, we discuss the late time  polynomial tails. In contrast to the ringdown, which is sensitive to the interior structure of the spacetime and depends on
the final attractor, the tails for the Yang-Mills field on an  asymptotically flat spacetime are universal. They decay as $s^{-4}$ in the interior ($|z|<1$) and as $s^{-2}$ along future null infinities ($z=\pm 1$) \cite{ch, bcr}. Moreover, the spatial profiles of tails also appear to be universal. This can be seen by the following heuristic  argument.  Consider a solution tending asymptotically to $W_0$. Substituting
\begin{equation}\label{profil-ansatz}
  W(s,z)=1+s^{-2} f(y), \quad \mbox{where}\quad y=s(1-z),
\end{equation}
into Eq.\eqref{eqz} and linearizing, we get
\begin{equation}\label{profil-eq}
  y(y+1) f'' + (2y-1) f' - 2f =0.
\end{equation}
Discarding the growing solution, we obtain
\begin{equation}\label{f}
  f(y)=\frac{A_R}{(1+y)^2}\,,
\end{equation}
hence for $0<z\leq 1$ and large $s$
\begin{equation}\label{prof_R}
  W(s,z)-1 \sim \frac{1}{s^2} \, \frac{A_R}{(1+s(1-z))^2}\,,
\end{equation}
where the amplitude $A_R$ is the only trace of initial data. By an analogous  argument, for $-1\leq z<0$ and large $s$
\begin{equation}\label{prof_L}
  W(s,z)-1 \sim \frac{1}{s^2} \, \frac{A_L}{(1+s(1+z))^2}\,.
\end{equation}
In general, $A_R\neq A_L$.
It follows from \eqref{prof_R} that for large $s$
\begin{equation}\label{deriv}
  \frac{\partial^n W}{\partial z^n} \bigg|_{z=1} \sim A_R (n+1)!\, s^{n-2},
\end{equation}
hence higher derivatives grow in time. This result can also  be derived by inserting the power series expansion near ${\mathcal J}_R^+$,
\begin{equation}
  W(s,z)=1+\sum_{n=0}c_n(s) (1- z)^n,
\end{equation}
 into Eq.\eqref{eqz} and solving the resulting system of ordinary differential equations iteratively starting from $c_0(s)=A_R s^{-2}$.
 The formula \eqref{deriv} reflects the fact that the decay of the Yang-Mills field is not uniform in space (because the decay  along ${\mathcal J}^+$ is slower than the decay in the interior) . By the conformal invariance of the problem at hand, the formula \eqref{deriv} holds for the derivatives of Yang-Mills field at the future horizon of the extremal Reissner-Nordstr\"om black hole. This type of instability for solutions of the wave equation on the extremal Reissner-Nordstr\"om black hole was discovered by Aretakis \cite{a1, a2} (see also \cite{bf} and \cite{lmrt} for a discussion of the relationship between the behavior of fields near the future (degenerate) horizon and the future null infinity).

The numerical evidence confirming the profile \eqref{prof_R} and the growth  of higher derivatives along the future null infinity \eqref{deriv} is shown in Figs.~6~and~7.

\begin{figure}[h]
  \begin{center}
    \includegraphics[width=0.75\textwidth]{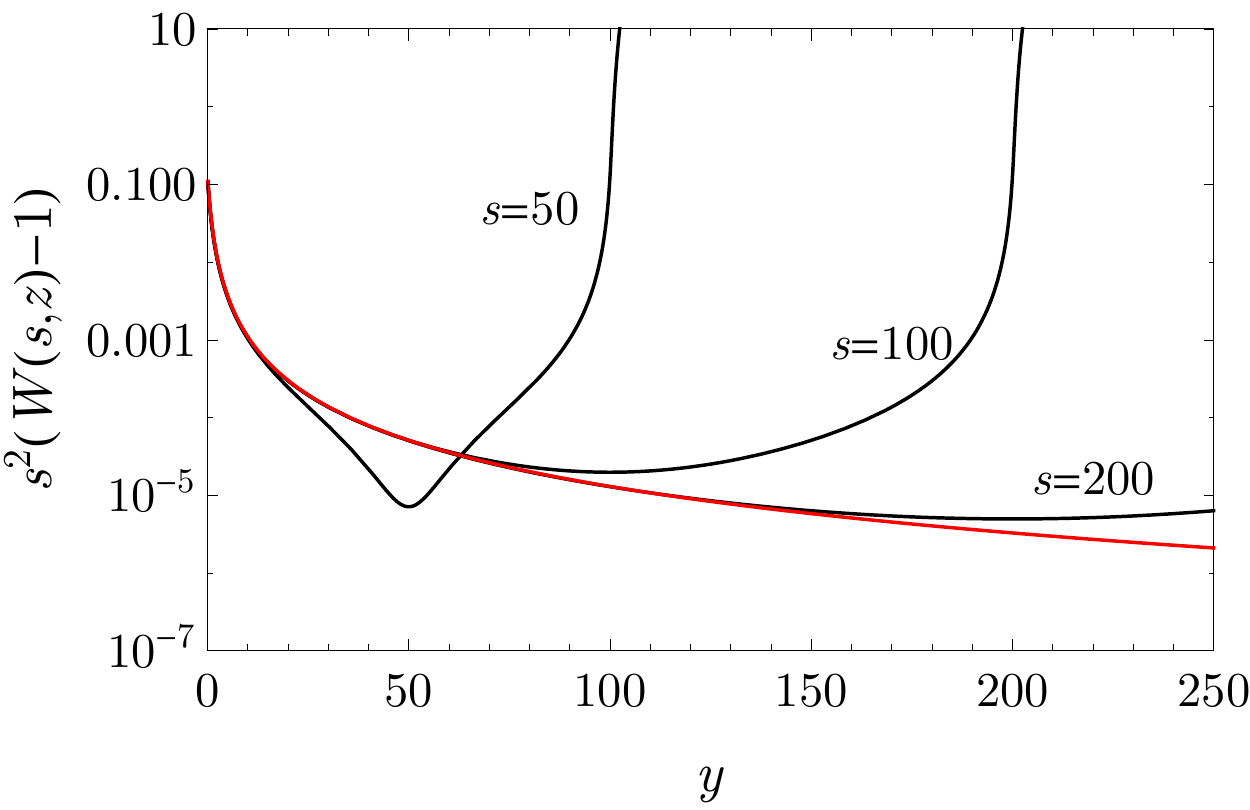}
    \caption{{\small Snaphots of the solution  shown in Fig.~3.  The late time profiles are seen to converge to the universal profile $f(y)$ (depicted in red) given by formula \eqref{f} with $A_R\approx 0.1319$.}}
    \label{fig7}
  \end{center}
\end{figure}

\begin{figure}[h]
  \begin{center}
    \includegraphics[width=0.75\textwidth]{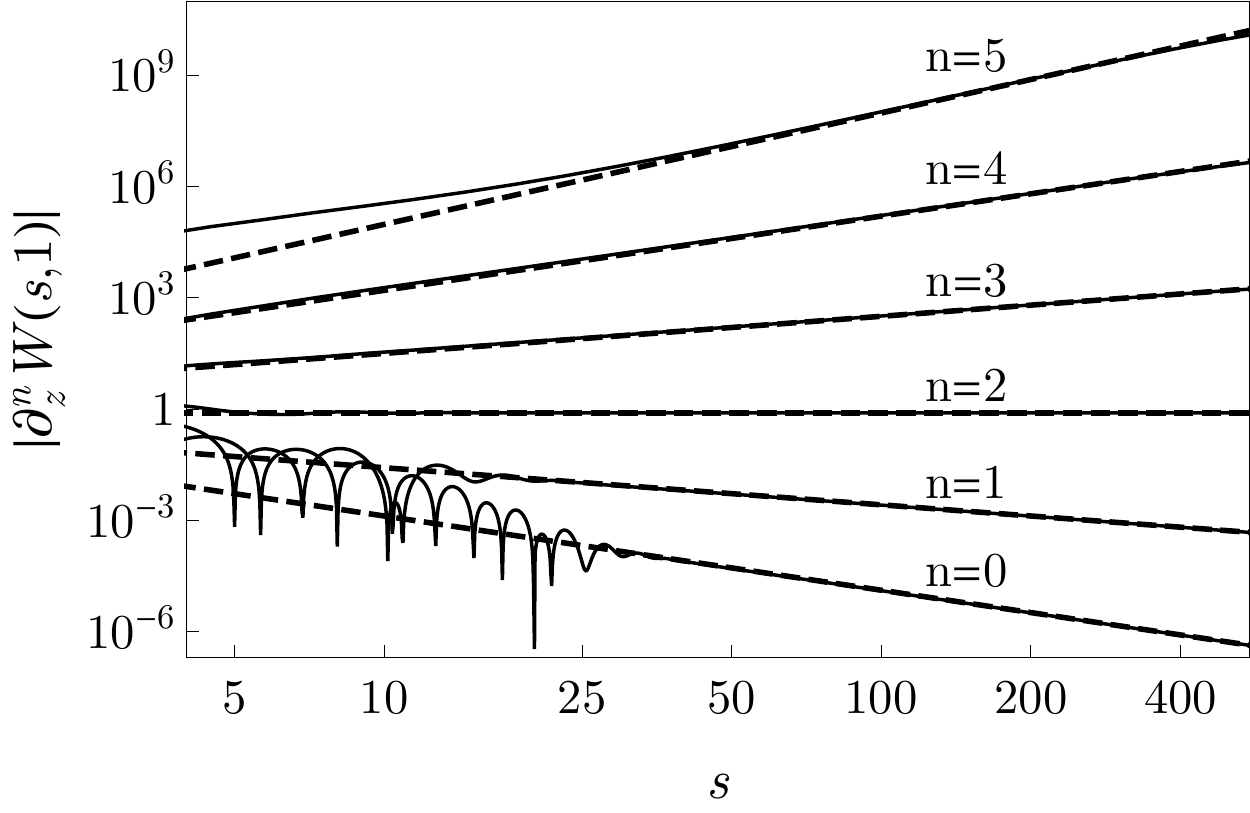}
    \caption{{\small For the solution shown in Fig.~3 we depict (in the log-log scale)  the transversal derivatives at ${\mathcal J}^+_R$.
    The late
     time  behaviour ($s>50$) agrees with  formula \eqref{deriv} for $A_R\approx 0.1319$.}}
    \label{fig6}
  \end{center}
\end{figure}

\subsubsection*{Acknowledgments.} We thank Pawe\l{} Biernat and Maciej Maliborski for helpful discussions.  This work was supported in part by the Polish National Science Centre grant no. DEC-2012/06/A/ST2/00397.
P.B. also gratefully acknowledges the support of the Alexander von Humboldt Foundation.

\end{document}